\documentclass[aps,showpacs,twocolumn]{revtex4}
\usepackage{amssymb}
\usepackage{graphicx}
\usepackage{dcolumn}
\usepackage{amsmath}

\begin{document}

\title{Anomalous ferromagnetism and non-Fermi-liquid behavior in the  Kondo lattice CeRuSi$_2$}
\author{V.N. Nikiforov$^1$, M. Baran$^2$, A. Jedrzejczak$^2$, V.Yu. Irkhin$^3$}
\email{Valentin.Irkhin@imp.uran.ru}
\affiliation{
$^1$Department of Physics, Moscow State University, 119899 Moscow, Russia}
\affiliation{
$^2$Institute of Physics, Polish Academy of Science, Lutnikov, Warsaw, Poland}
\affiliation{
$^3$Institute of Metal Physics, 620990 Ekaterinburg, Russia}

\begin{abstract}
The structural, electronic and magnetic properties of the Kondo-lattice system CeRuSi$_2$ are experimentally investigated and analyzed in the series of other ternary cerium compounds.
This system is shown to be an excellent model system demonstrating coexistence of the Kondo effect and anomalous ferromagnetism with a small magnetic moment which is confirmed by magnetic and $\mu$SR measurements.
Data on specific heat, resistivity, heat conductivity and Seebeck coefficient are presented.
Being deduced from the resistivity and  specific heat data, a non-Fermi-liquid behavior is observed at low temperatures, which is unusual for a ferromagnetic Kondo system. A comparison with other magnetic Kondo lattices is performed.

\end{abstract}

 \maketitle


\section{Introduction}
The cerium based ternary intermetallic compounds of CeTX$_2$ type (with T being transition metal and X semimetallic element like Si,  Ge  and  Sn) remain a subject of considerable interest because of their unusual ground state properties observed in many compounds of this family. For example, we have heavy-fermion behavior with a large electronic specific heat ($\gamma=$ 1.7 J/(mol K$^2$) at 1.25 K) in CePtSi$_2$ \cite{CePtSi}, valence fluctuation behavior in CeRhSi$_2$ \cite{CeRhSi} and in CeNiSi$_2$ \cite{CeNiSi}.
In the present work we discuss in detail the magnetic and electronic properties of the system CeRuSi$_2$ which has a ferromagnetic ground state \cite{CeRuSi}.

It was traditionally believed for many years that the competition of the intersite RKKY exchange interaction and the Kondo effect should result in the formation of either the usual magnetic ordering with large atomic magnetic moments (as in elemental rare-earth metals) or the non-magnetic Kondo state with suppressed magnetic moments \cite{Brandt}.
However, more recent experimental investigations have convincingly demonstrated that magnetic ordering and pronounced spin fluctuations are widely spread among heavy-fermion systems and other anomalous 4f- and 5f-compounds, which are treated usually as concentrated Kondo systems (see \cite{IK}).

The class of ``Kondo" magnets is characterized by (i) the logarithmic temperature dependence of resistivity typical for Kondo systems at not too low temperatures (ii) reduced value of the magnetic entropy at the ordering point, in comparison with the value $ R \ln (2J + 1)$ (which corresponds to usual magnets with localized moment $J$) (iii) small ordered magnetic moment $M_0$ (in comparison with the ``high-temperature" moment $\mu_{eff}$ determined from the Curie constant), which is reminiscent of weak itinerant magnets (iv) negative (even for ferromagnets) paramagnetic Curie temperature $\theta$ which strongly exceeds in absolute value the magnetic ordering temperature (this behavior is due to large single-site Kondo contribution to the paramagnetic susceptibility).

There exist numerous examples of antiferromagnetic systems where ``Kondo" anomalies in thermodynamic and transport properties coexist with magnetic ordering. At the same time, examples of Kondo ferromagnets are not so numerous: CeNiSb, CePdSb, CeSi$_x$, Ce$_2$Rh$_3$B$_2$, Sm$_3$Sb$_4$, NpAl$_2$ (the old bibliography is given in Ref.\cite{IK}).
The number of such materials gradually increases, including CePt \cite{CePt}, CeRu$_2$Ge$_2$ \cite{CeRuGe}, CeAgSb$_2$, \cite{CeAgSb},  CeRuPO \cite{CeRuPO}, CeRu$_2$M$_2$X (M = Al, Ga; X = B, C) \cite{CeRu2Ga2B1,CeRu2Ga2B}, CeIr$_2$B$_2$ \cite{CeIr2B2}, hydrogenated CeNiSn \cite{CeNiSn}.

In fact, a number of factors make the physical picture in the most ``Kondo" ferromagnets rather complicated. The systems like CeSi$_x$ \cite{613} are described by spin-fluctuation (rather than Kondo) model.
The Kondo systems YbNiSn \cite{YbNiSn} and UPdIn \cite{UPdIn} possess antiferromagnetic ordering and demonstrate canted ferromagnetism only.

In our opinion, CeRuSi$_2$ is a rather ``typical" anomalous rare-earth ferromagnetic system which exhibits the whole variety of peculiarities of the Kondo-lattice magnets. Moreover, it demonstrates the phenomenon of the non-Fermi-liquid behavior \cite{NFL} in low-temperature thermodynamic and transport properties.

\section{The  crystal structure}

We have investigated  properties of a large number of  CeTX$_2$, CeT$_2$X$_2$, Ce$_2$T$_3$X$_5$  and CeTX$_3$  compounds (with T = Ru, Rh, and X = Si, Ge). The high-purity samples were provided by Physical and Chemical Analysis Laboratory from the Chemical Department of the Moscow State University (guided by Yu. Seropegin). According to their crystal symmetry, these compounds can be divided into following classes: CeRuSi$_2$ (space group P21/m); CeRu$_2$Si$_2$, CeRu$_2$Ge$_2$, CeRh$_2$Si$_2$, CeRh$_2$Ge$_2$  (I4/mmm); CeRuSi$_3$, CeRuGe$_3$, CeRhSi$_3$ (I4mm); Ce$_2$Ru$_3$Ge$_5$, Ce$_2$Rh$_3$Si$_5$ (Ibam).

In the present work we focus on the most interesting compound  CeRuSi$_2$, although give in Sect. 4 a comparison with resistivity data for other systems mentioned.

The polycrystalline  samples  of  CeRuSi$_2$ were  synthesized  by  melting  the starting mixture  in  an  arc  furnace  in  an argon  atmosphere  followed  by  annealing  at  870K  for  600h. The purity of  the component metals was better  than 99.9\%.
Our  investigation  of  CeRuSi$_2$  \cite{CeRuSi} has shown  that  this  compound  crystallizes  in  a structure different from the CeNiSi$_2$  type which is common  for other compounds of  the CeTX$_2$ series.
The X-ray investigations  demonstrated that this compound has the NdRuSi$_2$-type low-symmetrical monoclinic crystal structure  \cite{CeRuSi}. The  sample material was  examined by  X-ray  analysis  and  identified  as  a  single CeRuSi$_2$ phase (the calculated X-ray density was approximately 9.485 g/cm$^3$).

The NdRuSi$_2$ structure  is  a  distortion  derivative  of  the orthorhombic CeNiSi$_2$-type structure characteristic for  other  CeTX$_2$  compounds.  A similar structure occurs for  LaRuSi$_2$ \cite{LaRuSi}.
The  unit  cell  parameters for CeRuSi$_2$ are: a = 4.478(1)\AA,   b = 4.093(1)\AA,   c = 8.302(5)\AA,  the  angle beta  being  equal 102.53(3) deg.  The  atomic   coordinates   are:
\newline
\newline
\noindent
\begin{tabular}{|c|cccc|}
\hline
Ce   &2(e)  &x/a = 0.4130(2)  &y/b = l/4   &z/c = 0.79904(9)\\
Ru   &2(e)  &x/a = 0.1179(2)  &y/b = l/4   &z/c = 0.3869(2)\\
Si1  &2(e)  &x/a = 0.0364(9)  &y/b = l/4   &z/c = 0.0907(5)\\
Si2  &2(e)  &x/a = 0.6657(9)  &y/b = l/4   &z/c = 0.4913(3)\\
\hline
\end{tabular}
\newline

The  CeRuSi$_2$ sample used in further measurements was of the weight 351.57 mg and of disk shape with the diameter of approximately 9 mm and thickness  1 mm.

\section{Magnetic properties}

The  magnetization  was  measured  by  a  SQUID  magnetometer  of
Quantum  Design  MPMS-5  in the broad interval of magnetic fields ($H<50$ kOe) and  temperatures  ($5<T<250$ K).
We will demonstrate that These data  give  evidence  for  the coexistence of  the Kondo  effect and  ferromagnetism in  CeRuSi$_2$.

\begin{figure}[htb]
\includegraphics[width=3.3in, angle=0] {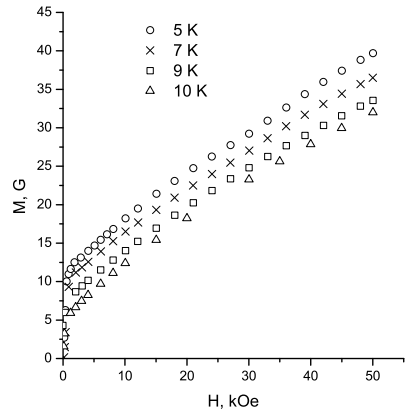}
\caption{The field dependences of magnetization at different temperatures}
\label{112MH}
\end{figure}

Fig. \ref{112MH} shows  the  field dependence  of  magnetization  $M(H)$  at  four temperatures.  The  $M(H)$  curve in the reversible  range  of  fields  (from  5  to  50 kOe)  are close to linear and are characterized by a relatively strong slope ($dM/dH$ is  about 5 10$^{-5}$ emu/g Oe)  that depends  slightly on temperature.
Saturation of the magnetic moment  in strong fields is absent (this is valid up to 150 T \cite{Arsamas}).

The $M(H)$ behavior in low fields  (rapid increase) is rather typical for ferromagnetic Kondo lattices like CeNiSb \cite{CeNiSb}, Ce$_{1-x}$La$_{x}$PdSb \cite{CePdSb}, CeRu$_2$(Si$_{1-x}$Ge$_{x}$)$_2$ \cite{CeRuSiGe}), Th$_{1-x}$U$_{x}$Cu$_{2}$Si$_{2}$ \cite{UCuSi2}.
Such an increase  is also reminiscent of ``itinerant metamagnetism" which occurs in a number of nearly or weakly ferromagnetic transition metal systems \cite{Levitin}.
A metamagnetic behaviour is  observed in the number of rare-earth compounds, e.g., for Ce$_5$Rh$_4$Sn$_{10}$ \cite{5410}, CeCoGa \cite{CeCoGa}. However, in our case there is no true metamagnetic transition, and ferromagnetism is retained in zero field, such a picture being  also not rare in itinerant ferromagnets (see Ref.\cite{Levitin}).
In this connection, it should be also mentioned that metamagnetism discussed earlier for CeRu$_2$Si$_2$ is  treated now as a continous pseudo-metamagnetic transition (crossover) \cite{Daou,Matsuda}, the itinerant character of 4f-electrons being preserved up to very high fields (about 50 T) \cite{Matsuda} which is similar to our system.

\begin{figure}[htb]
\includegraphics[width=3.3in, angle=0]{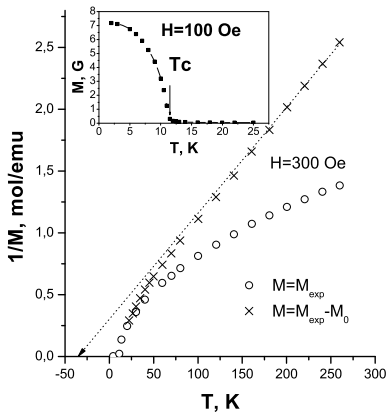}
\caption{The extraction of the Van Vleck contribution
The inset shows the  temperature  dependence of  magnetization at low temperatures
} \label{112VanVleck-mt}
\end{figure}

Now we discuss magnetic susceptibility.
Generally speaking, in the paramagnetic region (investigated up to 260 K) the simple Curie--Weiss law is not fulfilled (see temperature dependence of magnetization in small field, Fig. \ref{112VanVleck-mt}).
However, analyzing experimental data we were able to describe them by taking into account the Van Vleck contribution which is independent of temperature. Fitting $\chi = M/H = M_0/H + C_{eff}/(T - \theta)$ in the field $H = 300$ Oe to the experimental data, it was possible to get $M_0/H =$ 3.7 $10^{-6 }$ emu/g Oe, $C_{eff} =$ 1.22 10$^{-4}$ emu/g Oe and $\theta$  about $-40$ K. The value of $M_0/H$ is about 50\% of $M/H$ for 260K. This large Van Vleck contribution  is probably connected with low symmetry positions of Ce ions.

Thus the temperature dependence of magnetic susceptibility at temperatures above 60 K can be described by the Curie-Weiss law with the Curie constant $C_{eff}$, the corresponding value of the effective magnetic moment $\mu_{eff}$ being equal 1.7$\mu_B$ which is considerably smaller in comparison to that of the free Ce$^{3+}$ ion.

The paramagnetic Curie temperature $\theta$ is negative and large in absolute value. As discussed in the Introduction, this is typical for Kondo lattices (even ferromagnetic ones) since $\theta$ is determined by on-site Kondo screening rather than by intersite exchange interactions, $\chi(T = 0) \sim 1/T_K$.

The reduced value of the effective magnetic moment $\mu_{eff}$, calculated from the linear $\chi^{-1}(T)$ dependence, can be partly attributed to a moderate renormalization of the 4f shell moments due to hybridization with conduction electrons  as in Ce$_2$Rh$_3$B$_2$. A comparison with ferrimagnetic CeCoGa \cite{CeCoGa}  where $\theta = - 82$ K, $\mu_{eff}= 1.8\mu_B$ is also instructive.

At the same time, magnetic fluctuations in the presence of the Kondo effect can play a role.
Note that a similar strong reduction of effective moment is  observed near quantum critical point in Ce$_7$Ni$_3$ (under pressure) and Ce(Ru$_0.6$Rh$_0.4$)$_2$Si$_2$ (see the discussion in Ref.\cite{NFL}).





The inset in  Fig. \ref{112VanVleck-mt} shows the  temperature dependences  of  magnetization measured  in the field $H=100$ Oe. A  sharp upturn  of  $M(T)$ (as well as in the $\chi(T)$ curve)  below 11 K and  a  tendency  to  saturation  at  lowest temperature were detected.
Thus the  shape  of  the $M(T)$  and  $M(H)$ corresponds to the ferromagnet with the  transition temperature $T_c \simeq 11$ K.

\begin{figure}[htbp]
\includegraphics[width=3.3in, angle=0]{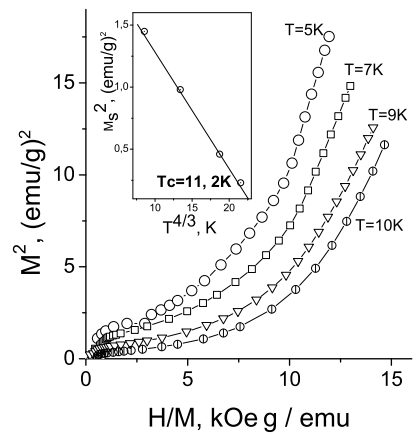}
\caption{Arrott plots of magnetization which determine the Curie temperature. Inset shows
the determination of $T_c$ from these plots} \label{Arrott}
\end{figure}

The fitted dependence $M(0)-M(T)$ below $T_c$ (in a wide temperature region, but at not too low $T$)
is  close to $T^{4/3}$ (see inset in Fig.\ref{Arrott}). Such a dependence is obtained in spin-fluctuation theories \cite{26} and is typical for weak itinerant ferromagnets where the spin-wave picture works at very  low $T$ only (as discussed in the Introduction, these systems have a number of similar features with the Kondo magnets).

\begin{figure}[htbp]
\includegraphics[width=3.3in, angle=0]{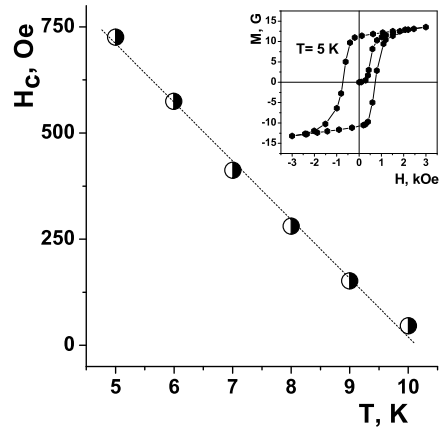}
\caption{The  temperature  dependence of  the coercive field. The inset shows hysteresis loop for $T = 5$ K }
\label{112HcT}
\end{figure}

We have found hysteresis loops for 5, 6, 7, 8, 9 and 10K (see inset in Fig. \ref{112HcT}).  From these loops we have estimated the coercive field $H_c$ as one half of loop width.

Fig. \ref{112HcT} shows the  temperature  dependence of $H_c(T)$ as  determined  from  the  hysteresis  loops
at  different  temperatures.



Taking the $T^{4/3}$  extrapolation of the $M(T)$ most steep part (between 10.5 and 11 K) to $x$-axis we should get $T_c=11.7$ K. From the above $M(H)$ data we have determined $M_s$ values  extrapolating linear dependence of $M$ in the $H=20-50$ kOe range to $H=0$.

The determination of the ferromagnetic transition temperature was also performed by the Belov-Arrott method. The field dependences of magnetization for $T =$ 5-10K are presented in Fig.\ref{Arrott} as the plots $M^2 = f(H/M)$.
Extrapolating $M^2$ to the value $H/M = 0$ enables one to obtain the value $M_s^2$, its temperature dependence yielding $T_c=$ 11.2 K.
The plot of $M_s^2(T)$ determining $ T_c$ is shown in inset of Fig. \ref{Arrott}.

The positive slope indicates  a second-order ferromagnetic phase transition, in agreement with the foregoing discussion. However, one can note that all the curves are not quite typical for usual ferromagnetic materials: they are non-linear and show downward curvature even in the high  field region. Somewhat better, but still not perfect linearity can be achieved by fitting the critical exponent $\beta \simeq 0.38$ in the modified Arrott plots ($M^{1/\beta} = f(H/M)$). This indicates a non-usual nature of ferromagnetism; a simple interpretation of  the non-linearity is presence of magnetic heterogeneity \cite{Aharoni}.


The relatively large slope in the reversible region of $M$ and the linearity in the $M(H)$ curves below the magnetic transition can be explained by the small value of the effective ground-state magnetic moment per Ce ion. The value $M=4$ emu/g obtained corresponds to $\mu_s =0.21\mu_B/$Ce ion.
This may give evidence for a complicated magnetic structure or, more likely, for a moment suppression by Kondo scattering which is responsible for the reduction of the $\mu_{eff}$ value at low temperatures.


To probe low-temperature magnetism, $\mu$SR investigations were also performed on the muon channel of the LNP
JINR phasotron using the spectrometer MUSPIN (some results were reported earlier \cite{Muon}).
Zero field (ZF), longitudinal field (LF) and transverse field (TF) $\mu$SR measurements have been carried out with a polycrystalline CeRuSi$_2$ sample.
The temperature interval was 4.2 K $<T<$ 300 K, above and below the Curie
temperature $T_c$. At all the temperatures,
the muon spin polarization function $P(t)$ has a non-oscillating form.
Below $T_c$ a hysteresis-like behavior of the polarization versus longitudinal magnetic field was found.

ZF data show a sharp increase of the muon relaxation rate below the temperature $T = 11.7$ K (0.42 $\mu$s$^{-1}$ at $T = 4.2$ K) justifying the phase transition to the magnetically ordered state. The results of LF measurements at $T = 4.2$ K show that the magnetic fields on the muon $B_\mu$, produced by the cerium magnetic moments, are mainly static: the external longitudinal field of 150 Oe practically recovers the muon spin polarization. In the paramagnetic phase the polarization decay has an exponential form at the temperatures 20 K $ <T<$ 300 K, although LF experiments at $T = 20$ K clearly show a significant static contribution (about 75\%). This situation can be discussed in the framework of a double relaxation model. The ferromagnetic type of magnetic ordering was proved by a hysteresis $B-H$ behavior observed in TF-experiment.

The lack of information about the muon site and type of magnetic ordering in
the compound does not allow the precise determination of the value of the ordered
moment. Nevertheless, taking into account the wide distribution of the magnetic field, we can estimate the value of the ordered magnetic moment about  0.05 $\mu_B$.

\section{Thermodynamic and transport properties}




\begin{figure}[htbp]
\includegraphics[width=3.3in, angle=0]{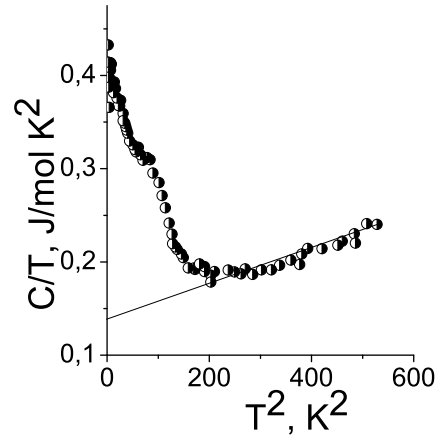}
\caption{The determination of $T$-linear electronic specific heat}
\label{gamma}
\end{figure}

The  low-temperature  $T$-linear electronic specific heat was obtained by the standard method, on using extrapolation from temperatures above 20 K, as shown in Fig. \ref{gamma}. The coefficient  $\gamma \simeq $ 140 mJ/mol K$^2$  is  markedly enhanced which means a moderately heavy-fermion behavior.

\begin{figure}[htbp]
\includegraphics[width=3.3in, angle=0]{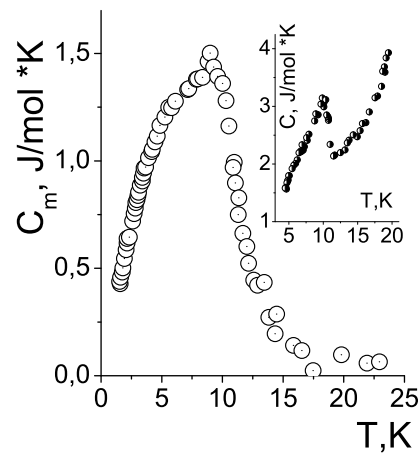}
\caption{The magnetic specific heat (after subtracting the $T$-linear and phonon contribution). The inset shows the temperature dependence of total specific heat }
\label{112Cmag}
\end{figure}

The specific heat anomaly at $T_c$ (Fig. \ref{112Cmag}) confirms the  presence  of  a magnetic  transition.
The entropy change near $T_c$ as calculated by integrating the $C(T)$ anomaly is relatively small: $\Delta S = 2.7$ J/K = 0.6 $R\ln2$ (where $R$ is the universal gas constant). The low value of  $\Delta S$ allows us to consider the 4f free-ion level of Ce to be split by a crystal electric field (CEF) with the doublet ($J= 1/2$) as a ground state. The high-temperature properties ($T \gg T_c$) are therefore determined by the combined action of Kondo effect and CEF splitting.

The results presented do not exclude a complex magnetic structure below 11K. Note that incommensurate magnetic ordering
is observed in a number of Kondo and heavy-fermion systems (see, e.g., \cite{CeAu2Ge2,CenMmIn3n+2m}).

\begin{figure}[htbp]
\includegraphics[width=3.3in, angle=0]{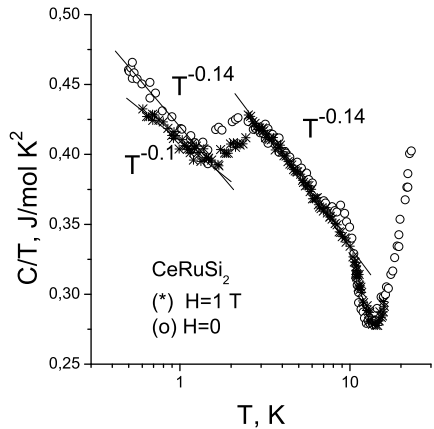}
\caption{The fit of temperature dependence of specific heat at ultralow temperatures in zero field and applied field $H=1$T}
\label{112CE}
\end{figure}

The fit of specific heat in the region of extremely low temperatures (below 2 K) demonstrates the logarithmic divergence  $C/T$ (J/mol K$^2$) = (3.25 - 1.08 $\log T$) 10$^{-4}$. This means a non-Fermi-liquid (NFL) behavior.
The temperature dependence can be also fitted by power-law dependences with small exponents (Fig. \ref{112CE}), a situation which is similar to many NFL systems \cite{NFL}.

One of the ways to govern the NFL behavior is applying the external magnetic filed \cite{Loehneysen}.
Fig. \ref{112CE} shows the influence of magnetic field $H=1$ T which suppresses somewhat the corresponding $C(T)$ anomaly (note that internal magnetic field in the ferromagnetic phase from $\mu$SR makes up 0.1 T only \cite{Muon}).
Unfortunately, we were not able perform the measurements in more strong fields which are required to clarify details of NFL behavior. Therefore further investigations are of interest.
Recently, the influence of magnetic field ion thermodynamic and transport properties of CeRu$_2$Si$_2$ was discussed in terms of Fermi surface reconstruction, which results in a plateau in the field dependence of specific heat \cite{Machida}.

Specific heat has also a weak anomaly near $2-3$ K  which is also influenced by magnetic field and may be related to crystal field effects. Similar peculiarities of NFL specific heat including the crystal field Schottky anomalies were observed in Ce$_{1-x}$Y$_{x}$RhIn$_{5}$ \cite{Zapf} and CePd$_{1-x}$Rh$_x$ \cite{Pikul}. In the latter system, demonstrating a transition
from ferromagnetism to intermediate valence state with $x$, the influence of magnetic field  on specific heat (suppression of NFL behavior) was  investigated too.

Again, Ce$_n$M$_m$In$_{3n+2m}$ heavy-fermion systems \cite{CenMmIn3n+2m} which demonstrate both incommensurate antiferromagnetic ordering and NFL properties (also in magnetic field) should be mentioned.

\begin{figure}[htbp]
\includegraphics[width=3.3in, angle=0]{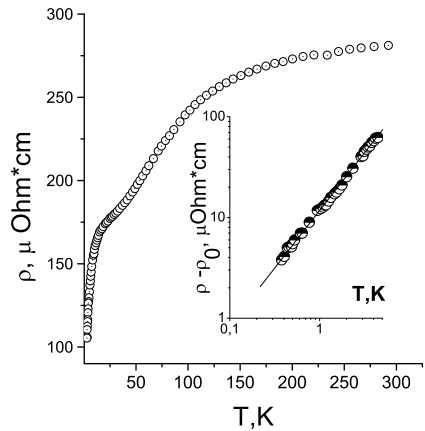}
\caption{The temperature dependence of  resistivity in a wide temperature region. The inset shows the resistivity in the low temperature region}
\label{resist}
\end{figure}

The  temperature  dependence  of  resistivity  $\rho (T)$ is shown in Fig.\ref{resist}). This exhibits  an anomaly  at the temperature $T_1 = T_c = 11.7$ K.

\begin{figure}[htbp]
\includegraphics[width=3.3in, angle=0]{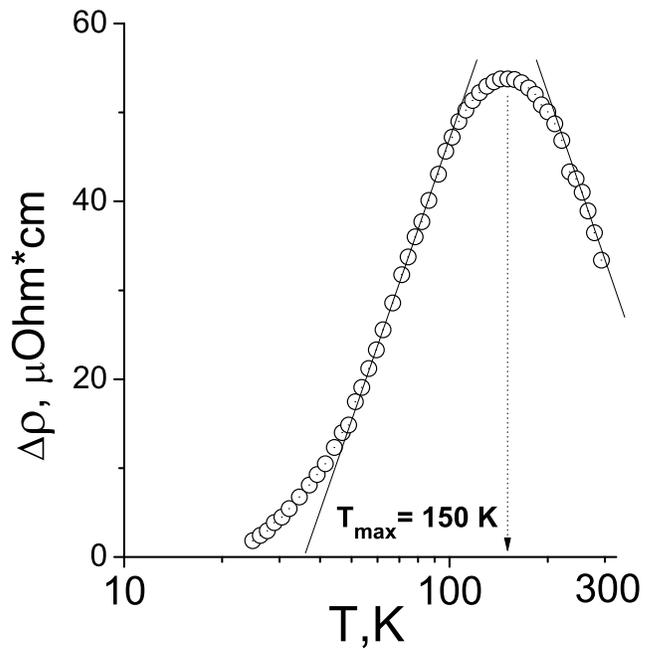}
\caption{The temperature dependence of  resistivity at high temperatures after subtracting the phonon contribution}
\label{112rtmax}
\end{figure}

The second anomaly (a maximum) occurs at high temperatures. To extract this, we eliminated the phonon contribution by using the Debye model.
The Debye temperature as determined from the $T^3$-contribution to specific heat (see Fig. \ref{gamma}) makes up $\theta_D = 226$ K.
After subtracting the corresponding Bloch-Gruneisen contribution we obtain the resistivity maximum at $T_2\simeq$ 150 K (Fig. \ref{112rtmax}).
Above $T_2$ we may pick up a high-temperature $T$-logarithmic Kondo contribution to resistivity, which is due to on-site $s-f$ scattering of conduction electrons by localized magnetic moments.
With decreasing temperatures, the lattice coherence effects result in resistivity decrease, which is typical for periodical Kondo systems. A role of crystal-field splitting effects is also possible in formation of the maximum at $T_2$.
Besides, in our system a rapid decrease of  $\rho (T)$  below  $T_1 = T_c$ is detected. We can propose that the magnetic transition at 11.7 K suppresses the Kondo scattering and leads to a decrease of the low-temperature resistivity.



It should be noted that for the widely investigated system CeRu$_2$Si$_2$ we have observed only one maximum ($T_2  = 50$ K) which has probably a Kondo origin.
The difference between CeRuSi$_2$ and CeRu$_2$Si$_2$ may be connected with a lower Kondo temperature in the latter system (its lattice cell volume is considerably larger).

On the other hand, for CeRu$_2$Ge$_2$ we have observed two resistivity anomalies at $T_1=8$ K and $T_2=$ 160 K. The first anomaly is accompanied by hysteresis and can be identified with a ferromagnetic transition, in agreement with the results of Ref.\cite{CeRuGe}. A magnetic transition was found also for CeRuGe$_3$ at $T=7$ K.

Following to the Sereni classification \cite{Sereni}, the difference in magnetic behavior of various (binary and ternary) cerium compounds may be attributed to different Ce--Ce distances.

In the low-temperature region, the resistivity of CeRuSi$_2$ obeys a non-Fermi-liquid (NFL) law, namely, $\rho (T) \propto T^\mu$ with $\mu = 1.1-1.2$ rather than a square dependence (inset in Fig.\ref{resist}). A similar behavior takes place in a number of anomalous rare-earth systems (see, e.g., \cite{CenMmIn3n+2m,NFL}).




\begin{figure}[htbp]
\includegraphics[width=3.3in, angle=0]{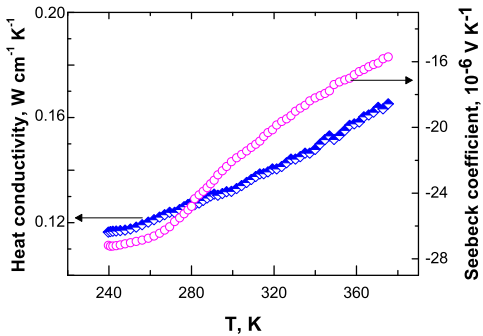}
\caption{The temperature dependence of  heat conductivity and thermoelectric power at high temperatures }
\label{seebeck}
\end{figure}

The Seebeck coefficient $\alpha$  measured at high temperatures (Fig.\ref{seebeck}) is rather high, temperature-independent region (or an extremum) being changed by a linear dependence. Such a behavior may confirm the presence of the Kondo scattering (cf. the discussion of thermoelectric power in the Kondo lattices in Ref. \cite{seebeck}). Observation of Kondo coherence features requires investigation at lower temperatures (cf., e.g., data for CeIr$_2$B$_2$ \cite{CeIr2B2}).
The corresponding figure of merit $ZT=\alpha^2T/\rho \kappa$ ($\kappa$ being heat conductivity) has a maximum at $T=250$ K where it makes up about 0.6\%.

\section{Conclusions and discussion}

To summarize, the results of the present paper and Ref.\cite{CeRuSi} allow us to conclude that CeRuSi$_2$
belongs to a relatively rare family of compounds where the Kondo effect and a ferromagnetism coexist.
The Kondo nature of ferromagnetism is confirmed by investigation of a wide range of electronic properties.
The shape of $M(H)$ and $M(T)$ curves, observation of hysteresis loops together with lambda-peak in the temperature dependence of specific heat $C(T)$ (see below) and the resistivity decrease are the characteristic features of a ferromagnetic transition at $T_1 = T_c =$ 11.7 K. An enhanced value of $\gamma \simeq 140$ mJ/mol K$^2$, reduced ground state moment, large absolute value of resistivity (being of hundreds $\mu$Ohm cm) together with observation of a maximum in $\rho(T)$ at $T_2 \simeq 150$ K, are characteristic features of dense Kondo effect. Thus, CeRuSi$_2$ might be considered as a ferromagnetic Kondo lattice.

Occurrence of non-collinear or inhomogeneous state near the magnetic transition is also possible. Neutron scattering experiments are required to investigate this issue.
At the same time, formation of a complicated  magnetic structure cannot be excluded, similar to the situation in YbNiSn (canted ferromagnetism \cite{YbNiSn}) or CeCoGa (ferrimagnetism \cite{CeCoGa}).
A possibility that Ru has a magnetic moment, as Co in CeCoGa, can be discussed.
On the other hand, neutron scattering analysis for CeRh$_3$B$_2$ reveals no magnetization on the rhodium and boron sites,  so that ferromagnetism originates  from the ordering of Ce local moments (and not, as has been claimed earlier, from itinerant magnetism in the Rh 4d band) \cite{Alonso}.

In our investigations, we did not exclude \emph{a priori} formation of spin-glass state. However,  our SQUID magnitometer measurements did not reveal an influence of prehistory (magnetization and demagnetization processes) on dc susceptibility. Thus within the accuracy of measurements FC and ZFC results coincided, in contrast with the spin-glass situation for Ce$_3$Pd$_{20}$Ge$_6$ where we detected   the ZFC-FC difference below 2.8 K \cite{Gaidukov} (such a spin-glass-like behavior was also found for CeCoGa \cite{CeCoGa} and CeRu$_2$(Si$_{1-x}$Ge$_{x}$)$_2$ \cite{CeRuSiGe}).
Further investigations of ac magnetic susceptibility would be useful.

Being deduced from the resistivity and specific heat data at ultralow $T$, the non-Fermi-liquid behavior is observed in our ferromagnetic Kondo compound.
To explain the NFL behavior in rare-earth and actinide systems, a number of mechanisms were proposed \cite{NFL}.  Since many known NFL systems are disordered alloys, one considers the influence of disordering in the Kondo-lattice model \cite{Miranda}, and the Griffiths singularity model \cite{Griffiths}.

In most Kondo magnets (see the Introduction) the NFl behavior can occur only on the boundary of magnetic instability, i.e. in the proximity of quantum critical point (quantum phase transition, QPT), which is governed by chemical composition  or external pressure (see, e.g., \cite{CeRu2Ga2B1}.

In this connection, one considers mechanisms connected with spin fluctuations \cite{26}, in particular treating the behavior near QPT in the clean limit \cite{Millis} or with account of disordering \cite{Belitz}.

In the system  Th$_{1-x}$U$_{x}$Cu$_{2}$Si$_{2}$ \cite{UCuSi2,NFL} NFL picture occurs in the  composition range  beyond where ferromagnetism is suppressed, so that $C/T$ behaves as $\ln T$ and the resistivity has linear $T$-dependence.
Further example is the URh$_{1-x}$Ru$_x$Ge alloy \cite{URhGe} where the critical concentration for the suppression of ferromagnetic order is $x_{cr}$ =
0.38. The Curie temperature vanishes linearly with $x$ and the ordered moment $\mu_s$ is suppressed in a continuous way ($\mu_s = 0.4 \mu_B$ for $x= 0$). At $x_{cr}$, the specific heat behaves as $T \ln T$, and the temperature exponent of the resistivity  attains a minimal value $\mu=1.2$. The total f-electron entropy obtained by integrating $C_{m}/T$ equals $0.48R\ln2$ for $x=0$ and decreases to $0.33R\ln2$ at $x_{cr}$.

In the system Ce$_{1-x}$La$_{x}$PdSb \cite{CePdSb}, demonstrating
high-temperature Kondo $\ln T$-contribution to  resistivity.
the ferromagnetic Kondo-lattice state transforms
gradually into the NFL state in the regime $x>0.7$, the Kondo disorder
model being a candidate for NFL mechanism.
The scaling of $\chi$, $\rho$ and $C/T$ with $T_K$ suggests that the NFL
behavior is a single ion phenomenon that is associated with
the Kondo effect.

Recently, a NFL picture somewhat similar to CeRuSi$_2$ has been reported for the ferromagnetic system URu$_{2-x}$Re$_x$Si$_2$ \cite{URu} which is characterized by a small ground state moment. In this system,
$C(T)=T \ln T$, $\rho(T) \propto T^\mu$ with $\mu$ = 1.2 over more than a decade in temperature below 20 K for $x = 0.6$.
The saturation moment, Curie temperature and electronic specific heat coefficient exhibit maxima for different Re concentration, reaching the values
$\mu_s = 0.44 \mu_B$/U atom, $T_c$= 38 K at x = 0.8, and $\gamma$ = 160 mJ/mol K$^2$ at $x =$ 0.6.
The ferromagnetic order was confirmed by neutron scattering experiments for $x = 0.8$ and NMR measurements for $x = 0.4$. However, corresponding magnetic-transition anomalies in specific heat and resistivity were not detected (unlike our investigations of CeRuSi$_2$).

The Griffiths-McCoy phase model \cite{Griffiths} was discussed in Ref.\cite{URu} as an explanation for the coexistence of ferromagnetism and NFL properties since the Griffiths singularities give rise to the NFL power-law behavior. In a clean system, spin-fluctuation mechanisms of NFL behavior, e.g., owing to peculiarities of magnetic ordering and spin dynamics in the Kondo lattices \cite{NFlIK}, are more probable.

The authors are grateful to Prof. H. Szymczak, Prof. R.Z. Levitin. R. Szymczak, A.V. Gribanov, and N.P. Danilova for useful discussions. This work is supported in part by the Programs of fundamental research of RAS Physical Division  ``Strongly correlated electrons in solids and structures", project No. 12-T-2-1001 (Ural Branch) and of RAS Presidium ``Quantum mesoscopic and disordered structures", project No. 12-P-2-1041.

\end{document}